\documentclass{WileyMSP-template}
\usepackage{amsmath}
\usepackage{graphicx}% Include figure files
\usepackage{dcolumn}% Align table columns on decimal point
\usepackage{bm}% bold math
\usepackage{braket}
\usepackage [latin1]{inputenc}

\def\bcen{\begin{center}}
\def\ecen{\end{center}}
\usepackage{epstopdf}
\begin{document}

\pagestyle{fancy}
\rhead{\includegraphics[width=2.5cm]{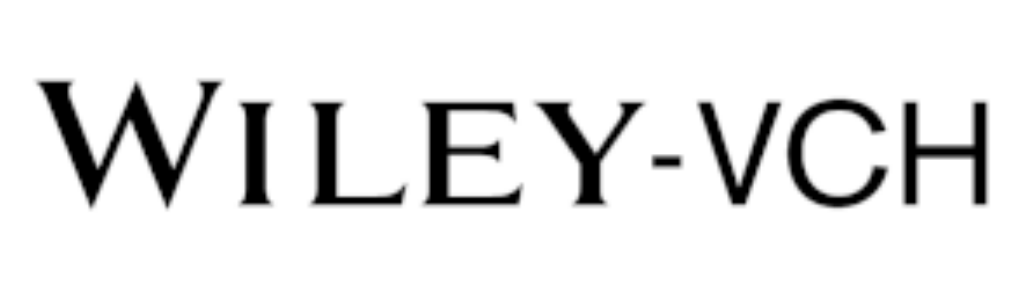}}

\title{Direct measurement of atomic entanglement via cavity photon statistics}

\maketitle

% Author: Please give full first and last names for authors and include * after the name of all corresponding authors

\author{Nilakantha Meher$^{1,3}$}
\author{M. Bhattacharya$^{2}$}
\author{Anand K. Jha$^{1}$}

% Dedication

%\dedication{Optional dedication here. If no dedication is required, please leave blank}

% Affiliations: Please provide adacemic titles (Prof. or Dr.) for all authors where applicable, and include an institutional email address for all corresponding authors
\begin{affiliations}
$^{1}$Department of Physics, Indian Institute of Technology Kanpur, Kanpur, UP 208016, India\\ $^{2}$School of Physics and Astronomy, Rochester Institute of Technology, 84 Lomb Memorial Drive, Rochester, New York 14623, USA\\
$^{3}$Current affiliation: Department of Chemical and Biological Physics,
Weizmann Institute of Science, Rehovot 7610001, Israel
\end{affiliations}

% Keywords: Please provide a minimum of three and a maximum of seven keywords, separated by commas


% Abstract should be written in the present tense and impersonal style (i.e., avoid we), and be at most 200 words long
\begin{abstract}
We propose an experimental scheme for the measurement of entanglement between two two-level atoms. Our scheme requires one of the two entangled atoms to interact with a cavity field dispersively, and we show that by measuring the zero time-delay second-order coherence function of the cavity field, one can measure the concurrence of an arbitrary Bell-like atomic two-qubit state. As our scheme requires only one of the atoms to interact with the measured cavity,  the entanglement quantification becomes independent of the location of the other atom. Therefore, our scheme can have important implications for entanglement quantification in distributed quantum systems.

\end{abstract}

\section{Introduction}
One of the most intriguing features of quantum mechanics is the quantum entanglement, which implies two or more quantum systems being inseparable even when they are space-like separated \cite{einstein1935pr,Bell1964PPF}. In the last several years, much effort has gone into generating and quantifying entanglement because of its wide applications in many quantum information protocols such as quantum teleportation \cite{bennett1993prl}, dense coding \cite{bennett1992prl}, etc. These protocols use two-qubit entangled states as the main resource and the amount of entanglement in these states decides the success probability of the protocols. Although for two-qubit systems there are several measures of entanglement, Wootter's concurrence is the most widely used one \cite{wootters1998prl}. Experimental measurement of concurrence of two-qubit states requires constructing the total density matrix through joint probability measurements \cite{White1999PRL}. However, for certain classes of two-qubit states, there are proposals for quantifying entanglement without reconstructing the entire density matrix \cite{Horodecki2002PRL,Horodecki2003PRL,Salles2006PRA,Romero2007PRA,
Lee2008PRA,Yang2009CTP,Zhang2013PRA,
Zhang2013PLA,Zhang2014EPJD}. Moreover, for pure two-qubit states, it has been shown that the reduced density matrix of one of the subsystems contains information about entanglement \cite{BengtssonBook}, implying that the two-qubit entanglement can be measured by making measurements on only one of the subsystems without requiring joint measurements \cite{Guhne2002PRA,Barbieri2003PRL,Walborn2006Nature,Pan2019PRL,
Meher2020JOSAB}. Implementing this concept, several groups have measured entanglement of such two-qubit states experimentally by making measurement on only one of the subsystems \cite{Barbieri2003PRL,Walborn2006Nature,Pan2019PRL}. This concept has also been used in the context of verifying entanglement in continuous-variable systems \cite{Jha2011PRA,Kulkarni2017NatComm,BhattacharjeeArxiv2020}. These techniques have several advantages as compared to the techniques that involve joint measurements \cite{Walborn2006Nature,Jha2011PRA}.

In the atomic domain, there are several methods for experimentally measuring the entanglement of pure two-qubit states based on joint measurement \cite{Romero2007PRA,Wilk2010PRL,Gaitan2010NJP,Hofmann2012Science,
Zhang2014EPJD}. There are also a few theoretical proposals that allow direct measurement of entanglement without implementing joint measurement \cite{Zhou2014PRA,Chen2017LPL}. These methods have their own advantages. However, the schemes for direct measurement of entanglement require two pairs of atoms to interact with four cavities, and one needs to control all the atoms simultaneously through individual quantum gate operations.  In this paper, we report a simpler scheme that allows direct quantification of entanglement between two two-level atoms. Our scheme requires one of the two entangled atoms to interact with a cavity field dispersively and the measurement of the zero time-delay second-order coherence function of the cavity field gives direct quantification of the concurrence. Using our scheme, one can measure the concurrence of arbitrary Bell-like atomic two-qubit states. 
Furthermore, the measurement of entanglement between two atoms becomes independent of the separation between the atoms as the measurement involves only one of the atoms. This type of measurement will have implications in quantifying entanglement in distributed quantum systems  \cite{Kimble2008Nature,Cacciapioti2020IEEE,Riebe2004Nature,Zheng2000PRL,
Barrett2004Nature,Bao2012PNAS,Xia2011JAP,Bose1999PRL,Arevalo2004JMO,
Schaetz2004PRL}.  
%The proposed scheme can have important implication for various protocols, for instance, quantum teleportation \cite{Riebe2004Nature,Zheng2000PRL,Barrett2004Nature,
%Bao2012PNAS,Xia2011JAP,Bose1999PRL}, quantum gates \cite{Serafini2006PRL,Zheng2009APL}, information swapping \cite{Arevalo2004JMO} and quantum dense-coding \cite{Schaetz2004PRL} in which atomic entanglement is used as main resource.

We have organized our article as follows: In Sec. \ref{ProposedSetup}, we have introduced our proposed experimental setup for the direct measurement of atomic entanglement. We present our methods for quantifying entanglement through cavity photon statistics in Sec. \ref{Methods}. We summarized our results in Sec. \ref{Summary}.
\section{Proposed experimental setup}\label{ProposedSetup}
\begin{figure*}
\begin{center}
\includegraphics[height=6.2cm,width=14cm]{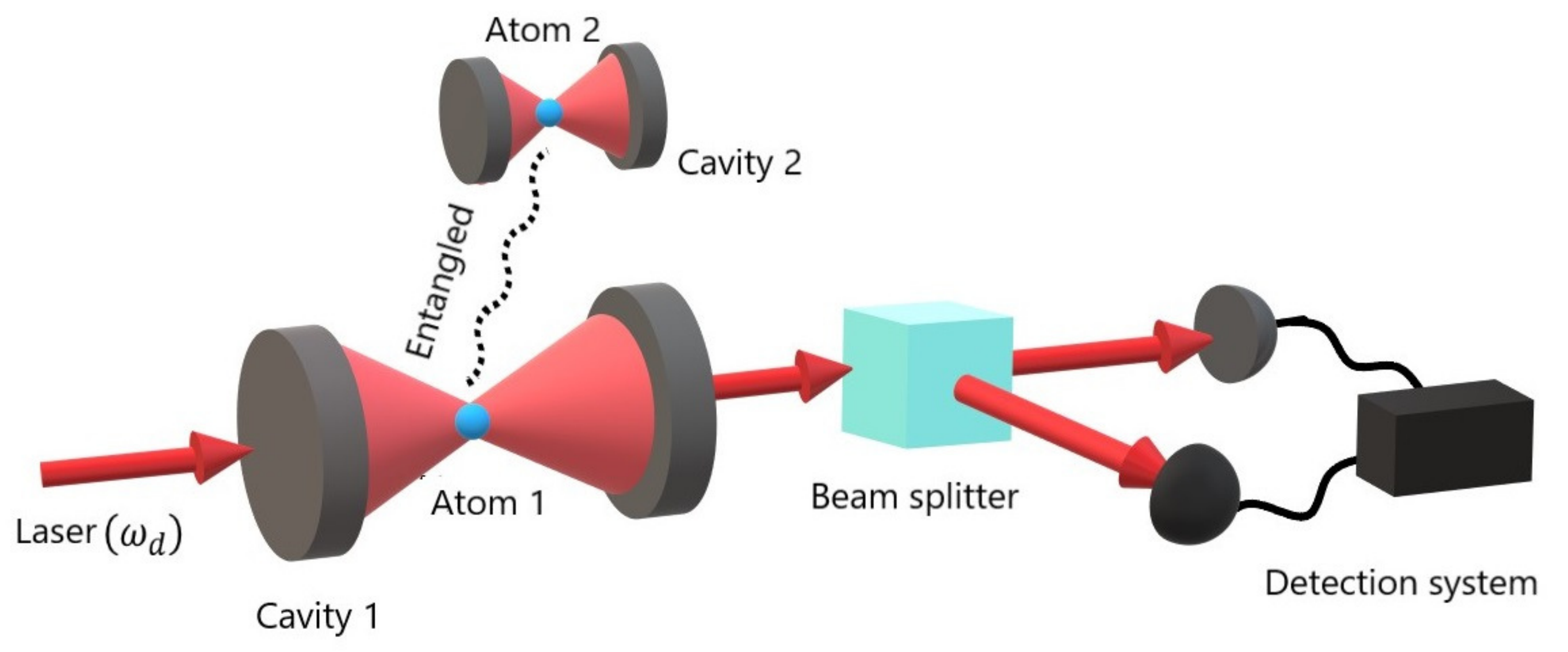}
\caption{Proposed experimental setup for quantifying entanglement between two two-level atoms through measurement of zero time-delay second-order coherence function $g^{(2)}(0)$ of one of the cavities. The setup consists of two spatially separated cavities (Cavity 1 and Cavity 2) each containing a two-level atom (Atom 1 and Atom 2). Cavity 1 is driven by a laser of resonance frequency $\omega_d$ whereas the Cavity 2 is in vacuum. The field leaks out from the driven cavity is directed towards the measurement system which consists of a beam splitter and a detection system for the purpose of measuring the zero time-delay second-order coherence function.}
\label{ExperimentalSetup}
\end{center}
\end{figure*}
We consider a system of two entangled two-level atoms (Atom 1 and Atom 2) which are trapped in two separated cavities, as shown in Fig. \ref{ExperimentalSetup}. These two atoms are interacting dispersively with their respective cavities. One of the cavities (Cavity 1) is driven by a laser with central frequency $\omega_d$ and the other cavity (Cavity 2) is in vacuum. As both the cavities are relatively separated, we write their Hamiltonians to be \cite{Brune1996PRL}
\begin{align}\label{mainHamiltonian}
H_1&=\frac{{\omega}_{a1}}{2}\sigma_{1}^z+(\Delta_r+\chi \sigma_{1}^z)a_1^\dagger a_1+\epsilon (a_1^\dagger +a_1),\nonumber\\
H_2&=\frac{{\omega}_{a2}}{2}\sigma_{2}^z+\omega_{c2}a_2^\dagger a_2+\chi \sigma_{2}^z a_2^\dagger a_2.
\end{align}
The Hamiltonian $H_1$ is written in the rotating frame of laser field. Here, $\Delta_r=\omega_{c1}-\omega_d$ is the detuning between the first cavity and the driving field, and $\epsilon$ is the driving strength.  $\omega_{aj}$ is the atomic transition frequency of $j$th atom and $\omega_{cj}$ is the cavity resonance frequency of $j$th cavity. The operator $\sigma_{j}^z=\ket{e_j}\bra{e_j}-\ket{g_j}\bra{g_j}$ is the atomic energy operator. We denote the excited state and ground state of $j$th atom by $\ket{e_j}$ and $\ket{g_j}$ respectively. The operator $a_j(a_j^\dagger)$ is the annihilation (creation) operator for $j$th cavity. The dispersive coupling strength is $\chi=g^2/\Delta$ \cite{Savage1990OptLett,Gerry}, where $g$ is the coupling strength between the first atom and first cavity, and the detuning between them is $\Delta=(\omega_{a1}-\omega_{c1})=(\omega_{a2}-\omega_{c2})$. 

The approximations considered in the Hamiltonians $H_1$ and $H_2$ are valid when the transition frequencies of the atoms are much larger than the resonance frequencies of the cavities, that is, $\omega_{a1}\gg\omega_{c1}$ and $\omega_{a2}\gg\omega_{c2}$. In this limit, the atom interacts with the cavity field dispersively and cannot exchange energy with the cavity \cite{Gerry1996PRA,Holland1991PRL}. However, it changes the phase of the cavity field \cite{Brune1996PRL,Holland1991PRL}. The dispersive coupling has been used for realizing quantum gates \cite{Heeres2015PRL}, generating nonclassical states \cite{Brune1992PRA,Govia2016PRA} and controlling photon transfer \cite{Nilakantha2020JOSAB,Meher2022Arxiv}, etc. The setup shown in Fig. \ref{ExperimentalSetup} consists of a beam splitter and a detection system for the purpose of measuring zero time-delay second-order coherence function of the field leaking out of the first cavity.

We further note that, as the second cavity is not driven, it  remains in vacuum. Thus, the coupling term $\chi \sigma_{2}^z a_2^\dagger a_2$ does not play any role in the dynamics. In experiment, one may not need the second cavity in the setup and it is the experimental choice to trap the second atom either in a cavity or in a trap \cite{Olmschenk2009Science}. However, we find that two separated cavities containing one atom in each forms a basic building for performing many quantum information tasks \cite{Cirac1997PRL,Davidovich1994PRA,Vogel2017QScTech}. Therefore, with this motivation, we consider such type of setup to present our analysis. 
\section{Measurement of entanglement}\label{Methods}
In this section, we present the scheme for direct measurement of entanglement between two atoms by measuring the zero time-delay second-order coherence function $g^{(2)}(0)$ of the field leaking out of the first cavity. Essentially, we connect the concurrence of atomic state to the $g^{(2)}(0)$ of the first cavity field. To calculate $g^{(2)}(0)$, we follow a master equation approach. In the presence of dissipation, under Born-Markov and rotating-wave-approximations, the dynamics of the system is described by the following master equation \cite{Carmichael}
\begin{align}\label{master}
\dot{\rho}&=-i[H_1+H_2,\rho]+\sum_{j=1}^2\frac{\kappa_j}{2}(2a_j\rho a_j^\dagger-a_j^\dagger a_j \rho-\rho a_j^\dagger a_j)\nonumber\\
&+\sum_{j=1}^2\frac{\gamma_j}{2}(2\sigma_{j}^-\rho \sigma_{j}^+-\sigma_{j}^+\sigma_{j}^-\rho-\rho\sigma_{j}^+\sigma_{j}^-),
\end{align}
where $\kappa_j$ is the decay rate of $j$th cavity and $\gamma_j$ is the atomic decay rate of $j$th atom.  We assume the cavity decay rates $(\kappa_1,\kappa_2)$ are much larger than the atomic decay rates $(\gamma_1,\gamma_2)$, such that the radiative life-time of atoms will be larger than the time required for the measurement of zero time-delay second-order coherence function of the cavity field. Our assumption is supported by the experimental observations, that is, the life-time of a superconducting qubit is several hundred times larger than the photon life-time inside a microwave cavity of low-quality factor \cite{Mlynek2014NatComm,Blais2004PRA,Yu2004PRL,Wallraff2005PRL}. Hence, to measure the entanglement from cavity spectrum before the atoms deteriorate, it is advantageous to choose the cavities having low-quality factor for our scheme. With this assumption, we neglect the atomic dissipation by setting $\gamma_1=\gamma_2=0$ in Eq. (\ref{master}) for deriving the zero time-delay second-order coherence function \cite{Guo2016PRA}. In the presence of cavity dissipation and driving, the cavity field reaches a steady state. 
%The time for reaching the steady state is infinitely large. However, if the cavity decay rates are larger, the expectation values of various observables will be very closed to their expected values in the steady state if $e^{-\kappa_1 t/2}$ is negligible. This provides an estimation of the time to reach to a nearly steady state on which the measurement can be performed experimentally.

\begin{figure}
\begin{center}
\includegraphics[height=7cm,width=12.5cm]{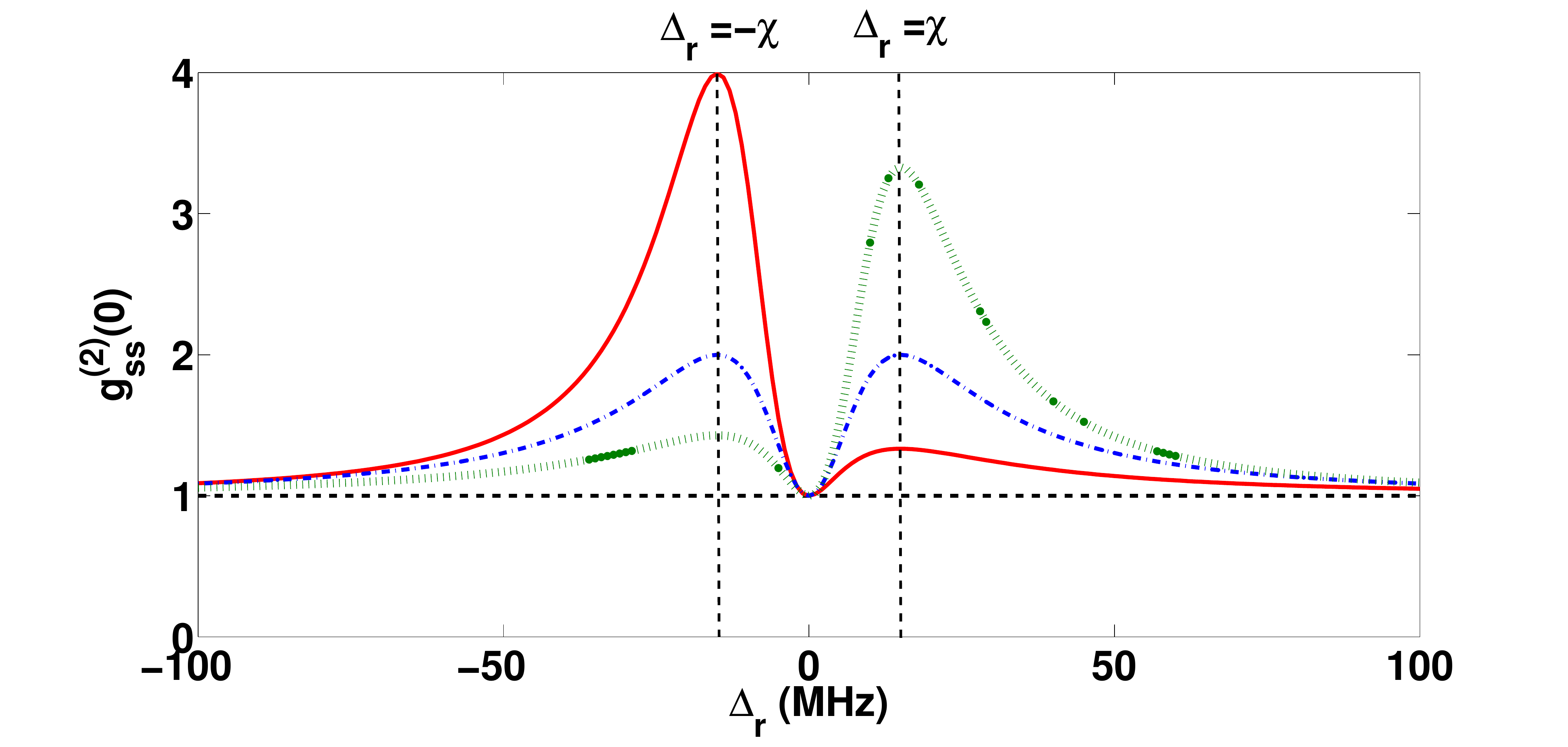}
\end{center}
\caption{$g_{\text{ss}}^{(2)}(0)$ as a function of $\Delta_r$ for the states with $\langle\sigma_{1}^z\rangle= 1$ (dashed), -0.5 (continuous), 0 (dot-dashed) and 0.4 (dotted). We use $\chi=15$ MHz and $\kappa_1=1$ MHz.}
\label{g2ssVsDrMain}
\end{figure}

The zero time-delay second-order coherence function for the cavity field in steady state is defined as \cite{LoudonBook,Mandel1965RevModPhys}
\begin{align}
g_{\text{ss}}^{(2)}(0)=\frac{\langle a_1^{\dagger 2}a_1^2 \rangle_{\text{ss}}}{\langle a_1^\dagger a_1\rangle_{\text{ss}}^2},
\end{align}
where $\langle a_1^{\dagger 2}a_1^2 \rangle_{\text{ss}}=\text{Tr}(a_1^{\dagger 2} a_1^2\rho_{\text{ss}})$ and $\langle a_1^\dagger a_1\rangle_{\text{ss}}=\text{Tr}(a_1^{\dagger} a_1\rho_{\text{ss}})$ are the steady state expectation values. Here $\rho_{\text{ss}}$ is the steady state density matrix in which the subscript \textquoteleft\textquoteleft ss\textquoteright\textquoteright~stands for steady state. These expectation values are to be calculated from a closed set of equations of motion by equating their time derivatives to zero. Using master equation given in Eq. (\ref{master}), we find the closed set of equations of motion to be
\begin{subequations} 
\begin{eqnarray}
&\frac{d\langle a_1^\dagger a_1 \rangle}{dt}=-i\epsilon(\langle a_1^\dagger \rangle-\langle a_1 \rangle)-\kappa_1\langle a_1^\dagger a_1 \rangle,\\
&\frac{d\langle a_1 \rangle}{dt}=-i\Delta_r\langle a_1 \rangle-i\chi\langle \sigma_{1}^z a_1 \rangle-i\epsilon-\frac{\kappa_1}{2}\langle a_1 \rangle,\\
&\frac{d\langle a_1^\dagger \rangle}{dt}=i\Delta_r\langle a_1^\dagger \rangle+i\chi\langle \sigma_{1}^z a_1^\dagger \rangle+i\epsilon-\frac{\kappa_1}{2}\langle a_1^\dagger \rangle,\\
&\frac{d\langle \sigma_{1}^z a_1 \rangle}{dt}=-i\Delta_r \langle \sigma_{1}^z a_1 \rangle-i\epsilon \langle \sigma_{1}^z \rangle-i\chi\langle a_1 \rangle-\frac{\kappa_1}{2}\langle \sigma_{1}^z a_1 \rangle,\\
&\frac{d\langle \sigma_{1}^z a_1^\dagger \rangle}{dt}=i\Delta_r \langle \sigma_{1}^z a_1^\dagger \rangle+i\epsilon \langle \sigma_{1}^z \rangle+i\chi\langle a_1^\dagger \rangle-\frac{\kappa_1}{2}\langle \sigma_{1}^z a_1^\dagger \rangle,\\
&\frac{d\langle \sigma_{1}^z \rangle}{dt}=0,\\
&\frac{d\langle a_1^{\dagger 2} a_1^2 \rangle}{dt}=2i\epsilon \langle a_1^{\dagger } a_1^2 \rangle-2i\epsilon \langle a_1^{\dagger 2} a_1 \rangle-2\kappa_1\langle a_1^{\dagger 2} a_1^2 \rangle,\\
&\frac{d\langle a_1^{\dagger 2} a_1 \rangle}{dt}=i\chi\langle a_1^{\dagger 2} a_1 \sigma_{1}^z\rangle+i\Delta_r \langle a_1^{\dagger 2} a_1 \rangle+2i\epsilon \langle a_1^{\dagger } a_1 \rangle-i\epsilon \langle a_1^{\dagger 2}  \rangle-\frac{3\kappa_1}{2}\langle a_1^{\dagger 2} a_1 \rangle,\\
&\frac{d\langle a_1^{\dagger } a_1^2 \rangle}{dt}=-i\chi\langle a_1^{\dagger } a_1^2 \sigma_{1}^z\rangle-i\Delta_r \langle a_1^{\dagger } a_1^2 \rangle-2i\epsilon \langle a_1^{\dagger } a_1 \rangle+i\epsilon \langle a_1^{ 2}  \rangle-\frac{3\kappa_1}{2}\langle a_1^{\dagger } a_1^2 \rangle,\\
&\frac{d\langle a_1^{\dagger 2} a_1\sigma_{1}^z \rangle}{dt}=i\chi\langle a_1^{\dagger 2} a_1 \rangle+i\Delta_r \langle a_1^{\dagger 2} a_1\sigma_{1}^z \rangle+2i\epsilon \langle a_1^{\dagger } a_1 \sigma_{1}^z\rangle-i\epsilon\langle a_1^{\dagger 2}\sigma_{1}^z \rangle-\frac{3\kappa_1}{2}\langle a_1^{\dagger 2 } a_1 \sigma_{1}^z\rangle,\\
&\frac{d\langle a_1^{\dagger } a_1\sigma_{1}^z \rangle}{dt}=-i\epsilon(\langle  a_1^\dagger \sigma_{1}^z \rangle-\langle a_1 \sigma_{1}^z \rangle)-\kappa_1\langle a_1^{\dagger } a_1\sigma_{1}^z \rangle,\\
&\frac{d\langle a_1^{\dagger 2 } \sigma_{1}^z \rangle}{dt}=2i\Delta_r \langle a_1^{\dagger 2} \sigma_{1}^z \rangle+2i\chi\langle a_1^{\dagger 2}  \rangle +2i\epsilon \langle a_1^{\dagger } \sigma_{1}^z \rangle-\kappa_1\langle a_1^{\dagger 2 } \sigma_{1}^z \rangle,\\
&\frac{d\langle a_1^{\dagger } a_1^2\sigma_{1}^z \rangle}{dt}=-i\chi\langle a_1^{\dagger } a_1^2 \rangle-i\Delta_r \langle a_1^{\dagger } a_1^2\sigma_{1}^z \rangle-2i\epsilon \langle a_1^{\dagger } a_1 \sigma_{1}^z\rangle+i\epsilon\langle a_1^{2}\sigma_{1}^z \rangle-\frac{3\kappa_1}{2}\langle a_1^{\dagger  } a_1^2 \sigma_{1}^z\rangle,\\
&\frac{d\langle a_1^{ 2 } \sigma_{1}^z \rangle}{dt}=-2i\Delta_r \langle a_1^{ 2} \sigma_{1}^z \rangle-2i\chi\langle a_1^{ 2}  \rangle -2i\epsilon \langle a_1 \sigma_{1}^z \rangle-\kappa_1\langle a_1^{2 } \sigma_{1}^z \rangle,\\
&\frac{d\langle a_1^{\dagger 2 }  \rangle}{dt}=2i\Delta_r\langle a_1^{\dagger 2 }  \rangle+2i\chi\langle a_1^{\dagger 2 }\sigma_{1}^z  \rangle+2i\epsilon\langle a_1^{\dagger  }  \rangle-\kappa_1\langle a_1^{\dagger 2 }  \rangle,\\
&\frac{d\langle a_1^{ 2 }  \rangle}{dt}=-2i\Delta_r\langle a_1^{ 2 }  \rangle-2i\chi\langle a_1^{ 2 }\sigma_{1}^z  \rangle-2i\epsilon\langle a_1  \rangle-\kappa_1\langle a_1^{ 2 }  \rangle.
\end{eqnarray}
\end{subequations} 
Note that we have not derived the equations of motion for the second atom-cavity system. Because the above set of equations is sufficient to obtain the zero time-delay second-order coherence function for the first cavity. We solve these equations for steady state, and we find the zero time-delay second-order coherence function for the first cavity to be
\begin{align}\label{g2ssmain}
g_{\text{ss}}^{(2)}(0)=\frac{(\frac{\kappa_1}{2})^4+A(\frac{\kappa_1}{2})^2+\chi^4+B\langle \sigma_{1}^z \rangle+6\chi^2 \Delta_r^2+\Delta_r^4}{\left(2\langle\sigma_{1}^z\rangle \chi \Delta_r-(\frac{\kappa_1}{2})^2-\chi^2-\Delta_r^2\right)^2},
\end{align}
where $A=2\chi^2-4\langle\sigma_{1}^z \rangle \chi \Delta_r+2\Delta_r^2$ and $B=-4\chi \Delta_r (\chi^2+\Delta_r^2)$. We note that $g_{\text{ss}}^{(2)}(0)$ depends on the state of the first atom via $\langle \sigma_{1}^z \rangle$. Here, $\langle\sigma_{1}^z\rangle$ has to be calculated using the state of the first atom. Hence, it is possible to control the photon statistics of the cavity field by changing the state of the first atom \cite{Guo2016PRA}. When the atom is not there in the cavity, that is, at $\chi=0$, we get $g_{\text{ss}}^{(2)}(0)=1$ indicating that the photon statistics of the cavity field is Poissonian. This is consistent with the fact that the cavity is driven by a laser and hence, one would expect Poissonian photon statistics of the cavity field. However, the presence of the atom changes the photon statistics of the cavity field. We show $g_{\text{ss}}^{(2)}(0)$ as a function of $\Delta_r$ for various values of $\langle\sigma_{1}^z\rangle$ in Fig. \ref{g2ssVsDrMain}.  As can be seen, $g_{\text{ss}}^{(2)}(0)=1$ for $\langle\sigma_{1}^z\rangle=\pm 1$ (dashed line). However, if $\langle\sigma_{1}^z\rangle\neq \pm 1$ then $g_{\text{ss}}^{(2)}(0)>1$ indicating the cavity field is super-Poissonian. We observe that $g_{\text{ss}}^{(2)}(0)$ has two peaks at $\Delta_r=\pm \chi$ in the large dispersive coupling limit, that is, $(1-\langle \sigma_{1}^z \rangle)\chi/\kappa_1\gg 1$. 
These two peaks have the values
\begin{align}\label{g2ss1-Sz}
g_{\text{ss}}^{(2)}(0)\left|\right._{\Delta_r=\chi} \approx \frac{2}{1-\langle \sigma_{1}^z\rangle},
\end{align}
for $\Delta_r=\chi$ and 
\begin{align}\label{g2ss1+Sz}
g_{\text{ss}}^{(2)}(0)\left|_{\Delta_r=-\chi}\right. \approx \frac{2}{1+\langle \sigma_{1}^z \rangle},
\end{align}
for $\Delta_r=-\chi$.

Now, we will show how entanglement is connected to the zero time-delay second-order coherence function $g_{\text{ss}}^{(2)}(0)$ of the cavity field. We consider the two atoms to be in Bell-like states, either in
\begin{subequations}\label{AtomicState}
\begin{eqnarray}
\ket{\Psi}=\alpha\ket{g_1}\ket{g_2}+\beta\ket{e_1}\ket{e_2},\\
\text{or}~~~~\ket{\Phi}=\gamma\ket{e_1}\ket{g_2}+\delta\ket{g_1}\ket{e_2},
\end{eqnarray}
\end{subequations}
where $\alpha,\beta,\gamma$ and $\delta$ are complex numbers satisfying $|\alpha|^2+|\beta|^2=1$ and $|\gamma|^2+|\delta|^2=1$. Here $\ket{e_1}\ket{g_2}$ represents the state of two atoms in which the first atom is in excited state and the second atom is in ground state, and so on. Bell-like atomic states have already been generated experimentally \cite{Hagley1997PRL,Wilk2010PRL,Matsukevich2006PRL} and have several important implications in various quantum information applications \cite{Raimond2001RevModPhys,Monroe2002Nature,Meher2020JPhyB}. First we will show how to measure the concurrence of the state $\ket{\Psi}$ from the zero time-delay second-order coherence function of the cavity field. For the state $\ket{\Psi}$, the expectation value $\langle \sigma_{1}^z\rangle=\text{Tr}(\rho_{a1}\sigma_{1}^z)=|\beta|^2-|\alpha|^2$, where $\rho_{a1}=\text{Tr}_{a2}(\ket{\Psi}\bra{\Psi})$ is the reduced density matrix of the first atom. Here $\text{Tr}_{a2}$ represents the partial trace over the state of the second atom. Now, multiplying Eqs.~(\ref{g2ss1-Sz}) and (\ref{g2ss1+Sz}), and putting the value of $\langle \sigma_{1}^z\rangle=|\beta|^2-|\alpha|^2$, we get
\begin{align}\label{g2D+g2D-}
g_{\text{ss}}^{(2)}(0)\left|\right._{\Delta_r=\chi} \times g_{\text{ss}}^{(2)}(0)\left|_{\Delta_r=-\chi}\right. &\approx \frac{4}{1-(|\beta|^2-|\alpha|^2)^2},\nonumber\\
&=\frac{4}{4|\alpha|^2|\beta|^2}.
\end{align}
We know from Wooter's concurrence formula \cite{wootters1998prl}, the concurrence for the state $\ket{\Psi}$ is $C=2|\alpha\beta|$. Then, putting this value of $C$ in Eq.~(\ref{g2D+g2D-}), we get  
\begin{align}\label{Concg2}
C \approx \frac{2}{\sqrt{g_{\text{ss}}^{(2)}(0)\left|\right._{\Delta_r=\chi} \times g_{\text{ss}}^{(2)}(0)\left|_{\Delta_r=-\chi}\right.}}. 
\end{align}
This is the central result of this article, which connects the zero time-delay second-order coherence function at two different values for the detuning $\Delta_r=\pm \chi$ with the concurrence of the atomic state. Hence, by measuring $g_{\text{ss}}^{(2)}(0)$ at two different values of detuning $\Delta_r$, one can directly quantify the entanglement between the atoms. A similar relation can be found between the concurrence and the zero time-delay second-order coherence function if the atoms are in the state $\ket{\Phi}$. However, we note that Eq.~(\ref{Concg2}) remains the same for the state $\ket{\Phi}$. This comes from the fact that the Wooter's concurrence for both $\ket{\Psi}$ and $\ket{\Phi}$ are same if they have equal superposition coefficients, that is, $\alpha=\gamma$ and $\beta=\delta$.
\begin{figure}
\begin{center}
\includegraphics[height=7cm,width=12.5cm]{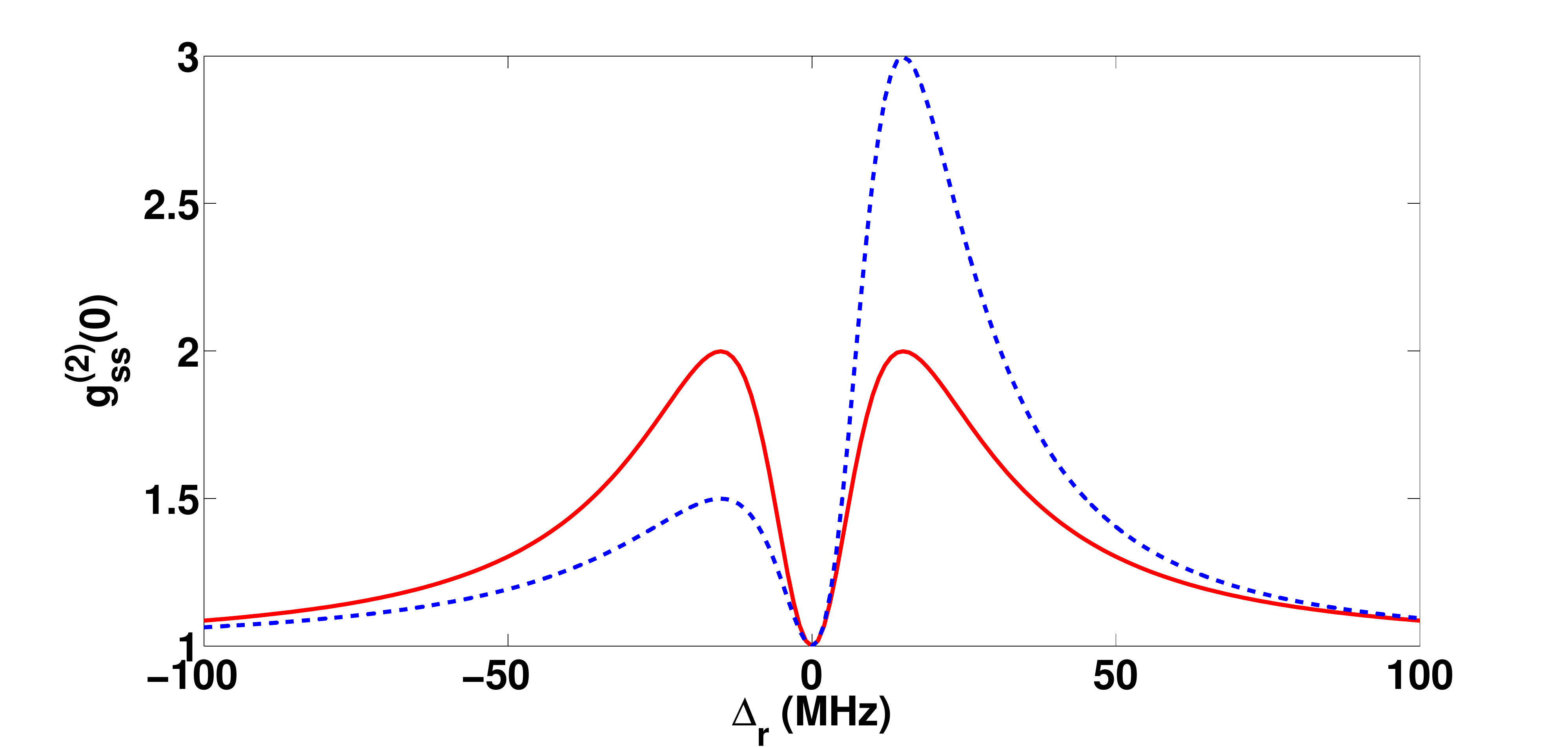}
\end{center}
\caption{$g_{\text{ss}}^{(2)}(0)$ as a function of $\Delta_r$ for the state $\ket{\Psi}=\frac{1}{\sqrt{2}}(\ket{g_1}\ket{g_2}+\ket{e_1}\ket{e_2})$ (continuous line) and $\ket{\Phi}=\sqrt{\frac{2}{3}}\ket{e_1}\ket{g_2}+\sqrt{\frac{1}{3}}\ket{g_1}\ket{e_2}$ (dashed line). We use $\chi=15$ MHz and $\kappa_1=1$ MHz.}
\label{g2ssVsDr}
\end{figure}

For the purpose of conceptual clarity, we consider two concrete examples:\\
\textbf{Example 1:} Let the state of the two atoms is $\ket{\Psi}=\frac{1}{\sqrt{2}}(\ket{g_1}\ket{g_2}+\ket{e_1}\ket{e_2})$. For this state, $\langle \sigma_{1}^z\rangle=0$. Using this value of $\langle \sigma_{1}^z\rangle$, we plot $g_{\text{ss}}^{(2)}(0)$ given in Eq.~(\ref{g2ssmain}) as a function of $\Delta_r$ in Fig. \ref{g2ssVsDr} for the state $\ket{\Psi}$ (continuous line) by taking the experimentally achievable values $\kappa_1=1$ MHz and $\chi=15$ MHz \cite{Mirhosseini2019Nature,Inomata2012PRB}. From the figure, we get $g_{\text{ss}}^{(2)}(0)\left|\right._{\Delta_r=\chi}=g_{\text{ss}}^{(2)}(0)\left|_{\Delta_r=-\chi}=2\right.$. Putting these values in Eq. (\ref{Concg2}), we get $C=1$, which is equal to the concurrence for the state $\ket{\Psi}$ that we get if we use Wooter's concurrence formula.\\
\textbf{Example 2:} Let the state of the two atoms be $\ket{\Phi}=\sqrt{\frac{2}{3}}\ket{e_1}\ket{g_2}+\sqrt{\frac{1}{3}}\ket{g_1}\ket{e_2}$. For this state, $\langle \sigma_{1}^z\rangle=1/3$. Using this value of $\langle \sigma_{1}^z\rangle$, we plot $g_{\text{ss}}^{(2)}(0)$ given in Eq.~(\ref{g2ssmain}) as a function of $\Delta_r$ in Fig. \ref{g2ssVsDr} for the state $\ket{\Phi}$ (dashed line). As can be seen in the figure, $g_{\text{ss}}^{(2)}(0)\left|\right._{\Delta_r=\chi}=3$ and $g_{\text{ss}}^{(2)}(0)\left|_{\Delta_r=-\chi}=1.5\right.$. Putting these values in Eq. (\ref{Concg2}), we get the concurrence for the state $\ket{\Phi}$ to be $C=2\sqrt{2}/3$, as expected.\\ 

The expression of $g_{\text{ss}}^{(2)}(0)$ given in Eq. (\ref{g2ssmain}) is derived by considering $\langle \sigma_{1}^z (t)\rangle=\langle \sigma_{1}^z (0)\rangle$ in the absence of atomic dissipation. If we consider atomic dissipation, the evolution equation of $\langle \sigma_{1}^z \rangle$ can be found from master equation to be ${d\langle \sigma_{1}^z \rangle}/{dt}=-\gamma_1 (\langle \sigma_{1}^z \rangle+1)$. The solution of this equation is $\langle \sigma_{1}^z(t) \rangle =e^{-\gamma_1 t}[\langle \sigma_{1}^z(0) \rangle+1]-1$, where $t$ is the time taken for completing the measurement. 
By choosing the atoms having long radiative life-time and cavities with low quality factor, one can minimize $\gamma_1 t$, such that $e^{-\gamma_1 t}\sim 1$ and $\langle \sigma_{1}^z(t) \rangle\approx \langle \sigma_{1}^z(0) \rangle$. Another factor that can minimize the $\gamma_1 t$ is the measurement time $t$ of $g^{(2)}(0)$. The measurement time $t$ should be much smaller than the radiative life-time of the atoms. We note that the life-time of a superconducting qubit, which is about 5 $\mu$sec ($\sim 0.2$ MHz) \cite{Mlynek2014NatComm}, is much larger than the time required for measuring $g^{(2)}(0)$, which is of the order of nanosecond, in superconducting cavity \cite{Grimm2019PRX,Rolland2019PRL,Hoi2012PRL}. The measurement time of $g^{(2)}(0)$ depends on the resolution time of the detector, which is usually of the order of nanosecond \cite{Grimm2019PRX,Rolland2019PRL,Hoi2012PRL,
Reinhard2012NatPhy,Faraon2008NatPhys}. The measurement time can be reduced further by using a detector having a smaller resolution time. However, recently, the second-order coherence function has been measured in the femtosecond range \cite{Boitier2009Nature}.  For a rough estimation, if we take the measurement time of $g^{(2)}(0)$ to be $t=50$ ns and the atomic decay rate to be $\gamma\sim 0.2$ MHz \cite{Mlynek2014NatComm}, then the exponential term will be $e^{-\gamma_1 t}\sim 0.99$ which is very close to unity and hence, our assumption $\langle \sigma_{1}^z (t)\rangle=\langle \sigma_{1}^z (0)\rangle$ is valid. Therefore, all the results derived in the previous section are robust even when the atomic decay is considered. We further note that our scheme requires to measure $g^{(2)}(0)$ at two different values of detuning $(\Delta=\pm \chi)$. After the measurement of $g^{(2)}(0)$ at a particular value of detuning, the measurement process may destroy the state of the atoms. Therefore, in experiment, one may need a few pairs of atoms being in the same entangled state to evaluate the concurrence from $g^{(2)}(0)$.    
\section{Conclusion and discussion} \label{Summary}
In summary, we have proposed a scheme for quantifying entanglement between two two-level atoms based on the measurement of zero time-delay second-order coherence function of a cavity field in which one of the atoms is dispersively interacting. Using our scheme, one can measure the concurrence of arbitrary Bell-like atomic two-qubit states.  Our scheme requires only one of the atoms to interact with the cavity field and the quantification does not depend on the location of the other atom. Hence, the entanglement quantification becomes independent of the separation between the atoms and this type of measurement can have implications in quantifying entanglement in distributed quantum systems \cite{Kimble2008Nature,Cacciapioti2020IEEE,Riebe2004Nature,Zheng2000PRL,
Barrett2004Nature,Bao2012PNAS,Xia2011JAP,Bose1999PRL,Arevalo2004JMO,
Schaetz2004PRL}. 
In addition, our scheme is based on detecting photons from the cavity which is less complicated than the measurement of atomic qubits and thus, it reduces the realization complexity. We also note that our scheme can be extended to other systems such as cavity-quantum dots \cite{Muller2007PRL,Hennessy2007Nature} and cavity-nitrogen vacancy centre \cite{Liu2013JAP,Berthel2015PRB} where the dispersive coupling is possible.

\section*{Acknowledgements}
We acknowledge financial support through the research
Grant No. DST/ICPS/QuST/Theme-1/2019 from the Department of of Science and Technology, Government of India.
\section*{Conflict of interest}
The authors declare that they have no conflict of interest.

%\bibliographystyle{MSP}
%\bibliography{EntanglemengtMeasurement}

\begin{thebibliography}{10}
\providecommand{\url}[1]{\texttt{#1}}
\providecommand{\urlprefix}{URL }

\bibitem{einstein1935pr}
A.~Einstein, B.~Podolsky, N.~Rosen,
\newblock \emph{Physical review} \textbf{1935}, \emph{47}, 10 777.

\bibitem{Bell1964PPF}
J.~S. Bell,
\newblock \emph{Physics Physique Fizika} \textbf{1964}, \emph{1} 195.

\bibitem{bennett1993prl}
C.~H. Bennett, G.~Brassard, C.~Cr\'epeau, R.~Jozsa, A.~Peres, W.~K. Wootters,
\newblock \emph{Phys. Rev. Lett.} \textbf{1993}, \emph{70}, 13 1895.

\bibitem{bennett1992prl}
C.~H. Bennett, S.~J. Wiesner,
\newblock \emph{Phys. Rev. Lett.} \textbf{1992}, \emph{69}, 20 2881.

\bibitem{wootters1998prl}
W.~K. Wootters,
\newblock \emph{Phys. Rev. Lett.} \textbf{1998}, \emph{80}, 10 2245.

\bibitem{White1999PRL}
A.~G. White, D.~F.~V. James, P.~H. Eberhard, P.~G. Kwiat,
\newblock \emph{Phys. Rev. Lett.} \textbf{1999}, \emph{83} 3103.

\bibitem{Horodecki2002PRL}
P.~Horodecki, A.~Ekert,
\newblock \emph{Phys. Rev. Lett.} \textbf{2002}, \emph{89} 127902.

\bibitem{Horodecki2003PRL}
P.~Horodecki,
\newblock \emph{Phys. Rev. Lett.} \textbf{2003}, \emph{90} 167901.

\bibitem{Salles2006PRA}
A.~Salles, F.~de~Melo, J.~C. Retamal, R.~L. de~Matos~Filho, N.~Zagury,
\newblock \emph{Phys. Rev. A} \textbf{2006}, \emph{74} 060303.

\bibitem{Romero2007PRA}
G.~Romero, C.~E. L\'opez, F.~Lastra, E.~Solano, J.~C. Retamal,
\newblock \emph{Phys. Rev. A} \textbf{2007}, \emph{75} 032303.

\bibitem{Lee2008PRA}
S.~M. Lee, S.-W. Ji, H.-W. Lee, M.~Suhail~Zubairy,
\newblock \emph{Phys. Rev. A} \textbf{2008}, \emph{77} 040301.

\bibitem{Yang2009CTP}
Y.~Rong-Can, L.~Xiu, H.~Zhi-Ping, L.~Hong-Cai,
\newblock \emph{Communications in Theoretical Physics} \textbf{2009},
  \emph{51}, 2 252.

\bibitem{Zhang2013PRA}
L.-H. Zhang, Q.~Yang, M.~Yang, W.~Song, Z.-L. Cao,
\newblock \emph{Phys. Rev. A} \textbf{2013}, \emph{88} 062342.

\bibitem{Zhang2013PLA}
L.-H. Zhang, M.~Yang, Z.-L. Cao,
\newblock \emph{Physics Letters A} \textbf{2013}, \emph{377}, 21 1421.

\bibitem{Zhang2014EPJD}
L.-H. Zhang, M.~Yang, Z.-L. Cao,
\newblock \emph{The European Physical Journal D} \textbf{2014}, \emph{68}, 5
  109.

\bibitem{BengtssonBook}
I.~Bengtsson, K.~Zyczkowski,
\newblock \emph{Geometry of Quantum States: An Introduction to Quantum
  Entanglement},
\newblock Cambridge University Press, \textbf{2006.}

\bibitem{Guhne2002PRA}
O.~G\"uhne, P.~Hyllus, D.~Bru\ss{}, A.~Ekert, M.~Lewenstein, C.~Macchiavello,
  A.~Sanpera,
\newblock \emph{Phys. Rev. A} \textbf{2002}, \emph{66} 062305.

\bibitem{Barbieri2003PRL}
M.~Barbieri, F.~De~Martini, G.~Di~Nepi, P.~Mataloni, G.~M. D'Ariano,
  C.~Macchiavello,
\newblock \emph{Phys. Rev. Lett.} \textbf{2003}, \emph{91} 227901.

\bibitem{Walborn2006Nature}
S.~P. Walborn, P.~H. Souto~Ribeiro, L.~Davidovich, F.~Mintert, A.~Buchleitner,
\newblock \emph{Nature} \textbf{2006}, \emph{440}, 7087 1022.

\bibitem{Pan2019PRL}
W.-W. Pan, X.-Y. Xu, Y.~Kedem, Q.-Q. Wang, Z.~Chen, M.~Jan, K.~Sun, J.-S. Xu,
  Y.-J. Han, C.-F. Li, G.-C. Guo,
\newblock \emph{Phys. Rev. Lett.} \textbf{2019}, \emph{123} 150402.

\bibitem{Meher2020JOSAB}
N.~Meher, A.~S.~M. Patoary, G.~Kulkarni, A.~K. Jha,
\newblock \emph{J. Opt. Soc. Am. B} \textbf{2020}, \emph{37}, 4 1224.

\bibitem{Jha2011PRA}
A.~K. Jha, G.~S. Agarwal, R.~W. Boyd,
\newblock \emph{Phys. Rev. A} \textbf{2011}, \emph{84} 063847.

\bibitem{Kulkarni2017NatComm}
G.~Kulkarni, R.~Sahu, O.~S. Magana-Loaiza, R.~W. Boyd, A.~K. Jha,
\newblock \emph{Nature Communications} \textbf{2017}, \emph{8}, 1 1054.

\bibitem{BhattacharjeeArxiv2020}
A.~Bhattacharjee, N.~Meher, A.~K. Jha,
\newblock \emph{arXiv:2102.04356 [quant-ph]} \textbf{2020}.

\bibitem{Wilk2010PRL}
T.~Wilk, A.~Ga\"etan, C.~Evellin, J.~Wolters, Y.~Miroshnychenko, P.~Grangier,
  A.~Browaeys,
\newblock \emph{Phys. Rev. Lett.} \textbf{2010}, \emph{104} 010502.

\bibitem{Gaitan2010NJP}
A.~Gaitan, C.~Evellin, J.~Wolters, P.~Grangier, T.~Wilk, A.~Browaeys,
\newblock \emph{New Journal of Physics} \textbf{2010}, \emph{12}, 6 065040.

\bibitem{Hofmann2012Science}
J.~Hofmann, M.~Krug, N.~Ortegel, L.~G{\'e}rard, M.~Weber, W.~Rosenfeld,
  H.~Weinfurter,
\newblock \emph{Science} \textbf{2012}, \emph{337}, 6090 72.

\bibitem{Zhou2014PRA}
L.~Zhou, Y.-B. Sheng,
\newblock \emph{Phys. Rev. A} \textbf{2014}, \emph{90} 024301.

\bibitem{Chen2017LPL}
L.~Chen, M.~Yang, L.-H. Zhang, Z.-L. Cao,
\newblock \emph{Laser Physics Letters} \textbf{2017}, \emph{14}, 11 115205.

\bibitem{Kimble2008Nature}
H.~J. Kimble,
\newblock \emph{Nature} \textbf{2008}, \emph{453}, 7198 1023.

\bibitem{Cacciapioti2020IEEE}
A.~S. {Cacciapuoti}, M.~{Caleffi}, F.~{Tafuri}, F.~S. {Cataliotti},
  S.~{Gherardini}, G.~{Bianchi},
\newblock \emph{IEEE Network} \textbf{2020}, \emph{34}, 1 137.

\bibitem{Riebe2004Nature}
M.~Riebe, H.~Haffner, C.~F. Roos, W.~Hansel, J.~Benhelm, G.~P.~T. Lancaster,
  T.~W. Korber, C.~Becher, F.~Schmidt-Kaler, D.~F.~V. James, R.~Blatt,
\newblock \emph{Nature} \textbf{2004}, \emph{429}, 6993 734.

\bibitem{Zheng2000PRL}
S.-B. Zheng, G.-C. Guo,
\newblock \emph{Phys. Rev. Lett.} \textbf{2000}, \emph{85} 2392.

\bibitem{Barrett2004Nature}
M.~D. Barrett, J.~Chiaverini, T.~Schaetz, J.~Britton, W.~M. Itano, J.~D. Jost,
  E.~Knill, C.~Langer, D.~Leibfried, R.~Ozeri, D.~J. Wineland,
\newblock \emph{Nature} \textbf{2004}, \emph{429}, 6993 737.

\bibitem{Bao2012PNAS}
X.-H. Bao, X.-F. Xu, C.-M. Li, Z.-S. Yuan, C.-Y. Lu, J.-W. Pan,
\newblock \emph{Proceedings of the National Academy of Sciences} \textbf{2012},
  \emph{109}, 50 20347.

\bibitem{Xia2011JAP}
Y.~Xia, J.~Song, P.-M. Lu, H.-S. Song,
\newblock \emph{Journal of Applied Physics} \textbf{2011}, \emph{109}, 10
  103111.

\bibitem{Bose1999PRL}
S.~Bose, P.~L. Knight, M.~B. Plenio, V.~Vedral,
\newblock \emph{Phys. Rev. Lett.} \textbf{1999}, \emph{83} 5158.

\bibitem{Arevalo2004JMO}
L.~M.~A. Aguilar, H.~Moya-cessa,
\newblock \emph{Journal of Modern Optics} \textbf{2004}, \emph{51}, 6-7 1089.

\bibitem{Schaetz2004PRL}
T.~Schaetz, M.~D. Barrett, D.~Leibfried, J.~Chiaverini, J.~Britton, W.~M.
  Itano, J.~D. Jost, C.~Langer, D.~J. Wineland,
\newblock \emph{Phys. Rev. Lett.} \textbf{2004}, \emph{93} 040505.

\bibitem{Brune1996PRL}
M.~Brune, E.~Hagley, J.~Dreyer, X.~Ma\^{\i}tre, A.~Maali, C.~Wunderlich, J.~M.
  Raimond, S.~Haroche,
\newblock \emph{Phys. Rev. Lett.} \textbf{1996}, \emph{77} 4887.

\bibitem{Savage1990OptLett}
C.~M. Savage, S.~L. Braunstein, D.~F. Walls,
\newblock \emph{Opt. Lett.} \textbf{1990}, \emph{15}, 11 628.

\bibitem{Gerry}
C.~Gerry, P.~Knight,
\newblock \emph{Introductory Quantum Optics},
\newblock Cambridge (England): Cambridge UP, \textbf{2005.}

\bibitem{Gerry1996PRA}
C.~C. Gerry,
\newblock \emph{Phys. Rev. A} \textbf{1996}, \emph{53} 2857.

\bibitem{Holland1991PRL}
M.~J. Holland, D.~F. Walls, P.~Zoller,
\newblock \emph{Phys. Rev. Lett.} \textbf{1991}, \emph{67} 1716.

\bibitem{Heeres2015PRL}
R.~W. Heeres, B.~Vlastakis, E.~Holland, S.~Krastanov, V.~V. Albert, L.~Frunzio,
  L.~Jiang, R.~J. Schoelkopf,
\newblock \emph{Phys. Rev. Lett.} \textbf{2015}, \emph{115} 137002.

\bibitem{Brune1992PRA}
M.~Brune, S.~Haroche, J.~M. Raimond, L.~Davidovich, N.~Zagury,
\newblock \emph{Phys. Rev. A} \textbf{1992}, \emph{45} 5193.

\bibitem{Govia2016PRA}
L.~C.~G. Govia, F.~K. Wilhelm,
\newblock \emph{Phys. Rev. A} \textbf{2016}, \emph{93} 012316.

\bibitem{Nilakantha2020JOSAB}
N.~Meher, S.~Sivakumar,
\newblock \emph{J. Opt. Soc. Am. B} \textbf{2020}, \emph{37}, 1 138.

\bibitem{Meher2022Arxiv}
N.~Meher, S.~Sivakumar,
\newblock \emph{arXiv:2204.01322} \textbf{2022}.

\bibitem{Olmschenk2009Science}
S.~Olmschenk, D.~N. Matsukevich, P.~Maunz, D.~Hayes, L.-M. Duan, C.~Monroe,
\newblock \emph{Science} \textbf{2009}, \emph{323}, 5913 486.

\bibitem{Cirac1997PRL}
J.~I. Cirac, P.~Zoller, H.~J. Kimble, H.~Mabuchi,
\newblock \emph{Phys. Rev. Lett.} \textbf{1997}, \emph{78} 3221.

\bibitem{Davidovich1994PRA}
L.~Davidovich, N.~Zagury, M.~Brune, J.~Raimond, S.~Haroche,
\newblock \emph{Phys. Rev. A} \textbf{1994}, \emph{50} R895.

\bibitem{Vogel2017QScTech}
B.~Vogell, B.~Vermersch, T.~E. Northup, B.~P. Lanyon, C.~A. Muschik
  \textbf{2017}, \emph{2}, 4 045003.

\bibitem{Carmichael}
H.~J. Carmichael,
\newblock \emph{{Statistical Methods in Quantum Optics 1: Master Equations and
  Fokker-Planck Equations}},
\newblock Springer, \textbf{1999}.

\bibitem{Mlynek2014NatComm}
J.~A. Mlynek, A.~A. Abdumalikov, C.~Eichler, A.~Wallraff,
\newblock \emph{Nature Communications} \textbf{2014}, \emph{5}, 1 5186.

\bibitem{Blais2004PRA}
A.~Blais, R.-S. Huang, A.~Wallraff, S.~M. Girvin, R.~J. Schoelkopf,
\newblock \emph{Phys. Rev. A} \textbf{2004}, \emph{69} 062320.

\bibitem{Yu2004PRL}
T.~Yu, J.~H. Eberly,
\newblock \emph{Phys. Rev. Lett.} \textbf{2004}, \emph{93} 140404.

\bibitem{Wallraff2005PRL}
A.~Wallraff, D.~I. Schuster, A.~Blais, L.~Frunzio, J.~Majer, M.~H. Devoret,
  S.~M. Girvin, R.~J. Schoelkopf,
\newblock \emph{Phys. Rev. Lett.} \textbf{2005}, \emph{95} 060501.

\bibitem{Guo2016PRA}
W.~Guo, Y.~Wang, L.~F. Wei,
\newblock \emph{Phys. Rev. A} \textbf{2016}, \emph{93} 043809.

\bibitem{LoudonBook}
R.~Loudon,
\newblock \emph{{The Quantum Theory of Light}},
\newblock Oxford University Press, \textbf{1983}.

\bibitem{Mandel1965RevModPhys}
L.~MANDEL, E.~WOLF,
\newblock \emph{Rev. Mod. Phys.} \textbf{1965}, \emph{37} 231.

\bibitem{Hagley1997PRL}
E.~Hagley, X.~Ma\^{\i}tre, G.~Nogues, C.~Wunderlich, M.~Brune, J.~M. Raimond,
  S.~Haroche,
\newblock \emph{Phys. Rev. Lett.} \textbf{1997}, \emph{79} 1.

\bibitem{Matsukevich2006PRL}
D.~N. Matsukevich, T.~Chaneli\`ere, S.~D. Jenkins, S.-Y. Lan, T.~A.~B. Kennedy,
  A.~Kuzmich,
\newblock \emph{Phys. Rev. Lett.} \textbf{2006}, \emph{96} 030405.

\bibitem{Raimond2001RevModPhys}
J.~M. Raimond, M.~Brune, S.~Haroche,
\newblock \emph{Rev. Mod. Phys.} \textbf{2001}, \emph{73} 565.

\bibitem{Monroe2002Nature}
C.~Monroe,
\newblock \emph{Nature} \textbf{2002}, \emph{416}, 6877 238.

\bibitem{Meher2020JPhyB}
N.~Meher,
\newblock \emph{Journal of Physics B: Atomic, Molecular and Optical Physics}
  \textbf{2020}, \emph{53}, 6 065502.

\bibitem{Mirhosseini2019Nature}
M.~Mirhosseini, E.~Kim, X.~Zhang, A.~Sipahigil, P.~B. Dieterle, A.~J. Keller,
  A.~Asenjo-Garcia, D.~E. Chang, O.~Painter,
\newblock \emph{Nature} \textbf{2019}, \emph{569}, 7758 692.

\bibitem{Inomata2012PRB}
K.~Inomata, T.~Yamamoto, P.-M. Billangeon, Y.~Nakamura, J.~S. Tsai,
\newblock \emph{Phys. Rev. B} \textbf{2012}, \emph{86} 140508.

\bibitem{Grimm2019PRX}
A.~Grimm, F.~Blanchet, R.~Albert, J.~Lepp\"akangas, S.~Jebari, D.~Hazra,
  F.~Gustavo, J.-L. Thomassin, E.~Dupont-Ferrier, F.~Portier, M.~Hofheinz,
\newblock \emph{Phys. Rev. X} \textbf{2019}, \emph{9} 021016.

\bibitem{Rolland2019PRL}
C.~Rolland, A.~Peugeot, S.~Dambach, M.~Westig, B.~Kubala, Y.~Mukharsky,
  C.~Altimiras, H.~le~Sueur, P.~Joyez, D.~Vion, P.~Roche, D.~Esteve,
  J.~Ankerhold, F.~Portier,
\newblock \emph{Phys. Rev. Lett.} \textbf{2019}, \emph{122} 186804.

\bibitem{Hoi2012PRL}
I.-C. Hoi, T.~Palomaki, J.~Lindkvist, G.~Johansson, P.~Delsing, C.~M. Wilson,
\newblock \emph{Phys. Rev. Lett.} \textbf{2012}, \emph{108} 263601.

\bibitem{Reinhard2012NatPhy}
A.~Reinhard, T.~Volz, M.~Winger, A.~Badolato, K.~J. Hennessy, E.~L. Hu,
  A.~Imamoglu,
\newblock \emph{Nature Photonics} \textbf{2012}, \emph{6}, 2 93.

\bibitem{Faraon2008NatPhys}
A.~Faraon, I.~Fushman, D.~Englund, N.~Stoltz, P.~Petroff, J.~Vuckovic,
\newblock \emph{Nature Physics} \textbf{2008}, \emph{4}, 11 859.

\bibitem{Boitier2009Nature}
F.~Boitier, A.~Godard, E.~Rosencher, C.~Fabre,
\newblock \emph{Nature Physics} \textbf{2009}, \emph{5}, 4 267.

\bibitem{Muller2007PRL}
A.~Muller, E.~B. Flagg, P.~Bianucci, X.~Y. Wang, D.~G. Deppe, W.~Ma, J.~Zhang,
  G.~J. Salamo, M.~Xiao, C.~K. Shih,
\newblock \emph{Phys. Rev. Lett.} \textbf{2007}, \emph{99} 187402.

\bibitem{Hennessy2007Nature}
K.~Hennessy, A.~Badolato, M.~Winger, D.~Gerace, M.~Atat�re, S.~Gulde,
  S.~F�lt, E.~L. Hu, A.~Imamoglu,
\newblock \emph{Nature} \textbf{2007}, \emph{445}, 7130 896.

\bibitem{Liu2013JAP}
S.~Liu, R.~Yu, J.~Li, Y.~Wu,
\newblock \emph{Journal of Applied Physics} \textbf{2013}, \emph{114}, 24
  244306.

\bibitem{Berthel2015PRB}
M.~Berthel, O.~Mollet, G.~Dantelle, T.~Gacoin, S.~Huant, A.~Drezet,
\newblock \emph{Phys. Rev. B} \textbf{2015}, \emph{91} 035308.

\end{thebibliography}

\end{document}